\documentclass[amssymb,aps,showpacs,amsmath,pra]{revtex4}

\usepackage{graphicx}
\usepackage{latexsym}

\newcommand{\be}{\begin{equation}}
\newcommand{\ee}{\end{equation}}

\newcommand{\bex}{\begin{eqnarray}}
\newcommand{\eex}{\end{eqnarray}}
\newcommand{\bmin}{\begin{center}\begin{minipage}{460pt}}
\newcommand{\emin}{\end{minipage}\end{center}}
\begin{document}

\title{Quantum seals} 
\author{Sudhir Kumar Singh}
\email{suds@ucla.edu}
\affiliation{Dept. of Electrical Engg., University of California, Los 
Angeles, CA 90095.}
\author{R. Srikanth}
\email{srik@rri.res.in}
\affiliation{Raman Research Institute, Bangalore- 560080, India.}


\begin{abstract}
A quantum seal is a way of encoding a message into quantum states,
so that anybody may read the message with little error, while authorized verifiers
can detect that the seal has been broken. We present a simple extension to the 
Bechmann-Pasquinucci majority-voting scheme that is impervious to coherent
attacks, and further, encompasses sealing quantum
messages by means of quantum encryption. The scheme is relatively
easy to implement, requiring neither entanglement nor controlled operations during
the state preparation, reading or verification stages.
\end{abstract}

\maketitle

\paragraph{Introduction:} Before the age of electronic communication, 
important documents were often closed
using a wafer of molten wax into which was pressed the distinctive seal of the sender.
This was meant to fulfil different purposes, namely authentication of the sender
as well as enabling the receiver to verify that the seal had not been broken,
and the message read, by a third party. Clearly it is meaningful to extend the
scheme to the digital world. Recently, Bechmann-Pasquinucci examined
a quantum scheme for sealing classical data \cite{bech}.
As with other related quantum cryptographic schemes,
such as quantum key distribution \cite{bb84,eke91,sp00,gis02}, 
it relies on the characteristic features
of quantum cryptology, namely 
the no-cloning theorem \cite{woo82} and quantum uncertainty to guarantee 
unconditional security whereas classical systems can offer at best computational
security. Specifically, Ref. \cite{bech} proposes a way to
represent  one bit of classical data by three qubits out of which one of them
(the seal qubit) is prepared in a diagonal basis state (an eigenstate of 
the Pauli $\hat{X}$ or $\hat{Y}$ operators), while the remaining
two (coding qubits) represent the classical bit in the computational basis (i.e., eigenbasis
of the Pauli $\hat{Z}$ operator).
Using single qubit measurement along the computational basis plus the classical
$[3,1,3]$-majority vote code, anyone can obtain the original classical bit with
certainty. And at the same time, the authorized verifier, who possesses some 
additional information on the seal qubit, is able to check if the seal was
broken with non-negligible probability. This scheme was extended in Ref. \cite{chau}
to the case of quantum messages, using quantum error correction codes.

In this article, we present a majority vote scheme that
guarantees an arbitrarily high probability that
a reader will unseal the correct message and is impervious even to 
coherent attacks of the type envisaged in Ref. \cite{bech}. It does not
require the application of any nonlocal (i.e., multi-qubit)
gate operations during the preparation, reading and verification stages 
and is hence fairly easy to implement. We believe that ease of implementability is of
significance, because if security were of prime concern, then quantum
key distribution, whose unconditional security has been extensively studied,
but which can be difficult and expensive to implement,
would be the appropriate way to protect communication. 

\paragraph{A modified scheme:} 
In our modified scheme, a classical bit is still read using a majority voting
system as in the Bechmann-Pasquinucci protocol,
but the coding bits are fewer than the sealing bits. That is, the fraction $f$
of coding qubits satisfies $f<1/2$. We use the notation where $\{|0\rangle, |1\rangle\}$
represent eigenstates of the computational basis, and $\{|\pm\rangle\}$ that of the
diagonal basis. For example, let us consider a 5-qubit seal (in practice, the seal must
be longer, as we note later) with two code bits, so that $f=2/5 = 0.4$.
A bit value $b$ can thus be encoded in any of the following
combinations, among others: 
\begin{eqnarray}
\label{listf}
|1\rangle|1\rangle|{+}\rangle|{-}\rangle|{+}\rangle \hspace{1.0cm}
{\rm code~bits~at~} 1,2; \hspace{1.0cm}b = 1\nonumber \\
|0\rangle|{+}\rangle|0\rangle|{-}\rangle|{+}\rangle\hspace{1.0cm}
{\rm code~bits~at~} 1,3; \hspace{1.0cm}b = 0\nonumber \\
|{-}\rangle|{+}\rangle|0\rangle|0\rangle|{+}\rangle\hspace{1.0cm}
{\rm code~bits~at~} 3,4; \hspace{1.0cm}b = 0\nonumber \\
|{+}\rangle|{-}\rangle|{+}\rangle|1\rangle|1\rangle\hspace{1.0cm}
{\rm code~bits~at~} 4,5; \hspace{1.0cm}b = 1
\end{eqnarray}
Further, the bit value encoded above is not a message bit: 
instead, each message bit is first split
into shares according to a secret sharing scheme \cite{schneier96}
and it is the share bits that are sealed and transmitted.
This ensures that the split-shared bit can be recovered only by combining the shares
in appropriate authorized combinations,
and as clarified below, improves security against illegal breaking of the seal.
In a simple instance, the shares could simply be $s$ single bits such that their bitwise
sum is the value of the message bit. 
Given $s$, a publicly known security parameter,
each message bit is classically split-shared into $s$ shares according to a
classical secret sharing scheme, in particular a $(s,s)$ threshold scheme.
Further protection can come by embedding
the share bit into the a classical error correction code \cite{coderate}.
It is therefore understood that the majority voting scheme described
below is applied not to the message
bit directly, but to the code bits derived from the message.

\paragraph{Encoding and verification by the sender:}
To seal a single code bit $b$, sender (sealer) Alice chooses $n$ qubits. 
Of these, a fraction $f < 0.5$ are prepared in the computational basis in the
state $|b\rangle$. 
These qubits are the code qubits. The remaining $(1-f)n$ qubits, which are the
seal qubits, are put randomly in any eigenstate of the diagonal basis 
$(1/\sqrt{2})(|0\rangle \pm |1\rangle)$.
First let us see that this suffices to ensure that, with high probability, anyone
can read the message, especially, considering that the code bits are not in
majority. By the large number theorem the expected number of seal bits
that if measured yield 0 or 1 is $(1-f)0.5n \pm \sqrt{(1-f)0.5n}$ (i.e., a square-root
statistical fluctuation).
The expected number of bits read as the intended, encoded bit is:
\be
\label{nfc}
nf_c \equiv n\left(f + \frac{1-f}{2}\right) = n\left(\frac{1}{2} + \frac{f}{2}\right).
\ee
To ensure that statistical fluctuation should not drown the signal, we will
require that $f$ should be sufficiently large: i.e., $fn > 
2\sqrt{(1-f)0.5n}$ or $n > n_0 \equiv 2(1-f)/f^2$. Thus, for example, if we choose $f=0.4$,
the sealed message length should be greater than 8 qubits. If we choose the more
delicate $f=0.25$, then the sealed message length should be greater than 24 qubits.
Conversely, given $n$, $f > f_0 \equiv (\sqrt{2n+1}-1)/n$. For example, if $n=40$ qubits,
then $f_0 = 0.2$, so that the code qubits should be more than 8.

This would seem to suggest that the larger is $f$, the better. This is indeed true for
plainly breaking the seal and reading the message. However, it also increases insecurity of a kind:
suppose one randomly picks one qubit and measures its state in the 
computational basis. The chance of knowing the message bit without being caught equals
$p_{\rm cheat} = f + (1-f)(1/2)(1/2) = (3/4)f+1/4$. 
In the scheme of Ref. \cite{bech}, $f=2/3$, so that $p_{\rm cheat}=3/4$. 
By setting $f\rightarrow0$, we obtain the limit
cheating probability of this kind to be (1/4). In practice, in order to
guard against statistical flucatuations, $f$ should be chosen greater than $f_0$.
Even a cheating probability of $1/4$ is too large. However, because of the
secret sharing the probability that the cheater is caught rapidly rises according to
$(1-p_{\rm cheat}^s)$.

Suppose the message has been read by someone who measures all qubits in the
computational basis.
In order to verify whether the seal is broken, Alice measures $r \le (1-f)n$
seal qubits on known coordinates in the diagonal basis. She checks that the 
outcomes match her preparation record. With high probability $(= 1 - (1/2)^r)$
she will detect at least one mismatch if the seal has been broken, and thus know
that the message has been read. (In this work, we ignore the effect of noise).
Notice that after reading the message, a reader on average knows $n(1-f)/2$
coordinates, where the minority outcome was obtained, to be the seal qubits. But
he learns nothing of their original state because of the no-cloning theorem
and quantum uncertainty.

\paragraph{Intended reader verification:}
In the case of a classical seal, the receiver is familiar with the design 
of the symbol pressed into the wax, and uses this knowledge to identify the
document as authentic. This means prior knowledge on the part of the intended receiver
or verifier is required. For the present
quantum seal, this part of the protocol is obtained by Alice providing 
a distinct set of $r_j < (1-f)n$ coordinates of the seal qubits and the corresponding 
preparation information to each authorized reader. The authorized verifier 
uses projective measurements to determine that the seal qubits have not been
disturbed, leaving
the code qubits untouched. He himself cannot read it because he has
information only on part of the seal qubits, and does not know which of the remaining
qubits are code qubits and which seal qubits. Let $R_j$ be the set of coordinates Alice
gives to the $j$th authorized verifier along with the corresponding preparation
information. In order that all verifiers should not be
able to collude and read the message without breaking the seal, we require that
the union of their sets should be a proper subset of the seal qubits. As a consequence,
we have $|\cap_j R_j| < (1-f)n$. This will demand a sufficiently large number of
seal qubits. An alternative scheme is to give the verifiers coordinate information,
and quantum information of the seal qubit states,
for performance of a non-desctructive state comparison
involving a control-swap gate \cite{bech}. 
From the viewpoint of implementation, these multi-qubit operations are more difficult
relative to the plain projective measurements in our case.

\paragraph{Security aspects:}
As pointed out in Ref. \cite{bech}, a majority encoding scheme (with $f > 0.5$) 
will necessarily be insecure against a coherent attack,
i.e., one based on collective, incomplete measurement or 
a suitable positive operator-valued measure (POVM) on all qubits taken together.
The reason is that any encoding for a bit is orthogonal to every encoding for the
other bit. For example, in a 3-qubit seal with $f=2/3$, the subspace of all states
that can encode for $b=0$ is spanned by the vectors $\{|0\rangle|0\rangle|0\rangle,
|0\rangle|0\rangle|1\rangle,|0\rangle|1\rangle|0\rangle,|1\rangle|0\rangle|0\rangle\}$,
whereas that of states encoding for $b=1$ by 
$\{|1\rangle|1\rangle|1\rangle,
|1\rangle|1\rangle|0\rangle,|1\rangle|0\rangle|1\rangle,|0\rangle|1\rangle|1\rangle\}$.
Since these two subspaces are mutually orthogonal, an incomplete three-qubit measurement 
can in principle distinguish them \cite{bech}.

In general, let $\rho_b$ denote the mixed state encoding
for code bit $b$ as seen by a potential attacker. 
A scheme where $\rho_0$ and $\rho_1$ have mutually orthogonal support is insecure
towards a coherent attack. An important feature of our scheme is that
because $f < 0.5$, the supports for $\rho_0$ and $\rho_1$ are not mutually orthogonal.
For example, in the partial listing (\ref{listf}), the first and third vectors
are not orthogonal to each other even though they encode for complementary bits;
likewise nor are the second and fourth vectors.
Indeed, any valid encoding for a bit value (say 0)
will be non-orthogonal to $^{(1-f)n}C_{fn}2^{fn}$
valid encodings for the other bit value (in this case, 1). 
Here we note that including the mutually unbiased basis states of the Pauli $\hat{Y}$ operator
improves the above count to  $^{(1-f)n}C_{fn}4^{fn}3^{(1 - 2f)n}$. However, the 
original, simpler scheme also suffices to guarantee unconditional (i.e., exponential in
some security parameter) security.
To see that as a result the two density operators approach indistinguishability rapidly,
we use a simple measure of closeness between $\rho_0$ and $\rho_1$,
the Hilbert-Schmidt distance:
\be
\label{dhs}
d_{\rm HS}^2 = {\rm Tr}\left[(\rho_0 - \rho_1)^{\dag}(\rho_0 - \rho_1)\right].
\ee
To a potential quantum attacker who only knows $f$, the state encoding for a 0 bit
is $\rho_0 = \left(^{n}C_{fn}2^{(1-f)n}\right)^{-1}(|0\rangle\langle0|\otimes
|0\rangle\langle0|\cdots|0\rangle\langle0|\otimes|s_1\rangle\langle s_1|
\otimes\cdots\otimes|s_{(1-f)n}\rangle\langle s_{(1-f)n}| + \cdots)$
where the summation runs over all $^{n}C_{fn}$ combinations for interspersing
the ${fn}$ copies of the 
$|0\rangle$ bits amidst the remaining $(1-f)n$ seal qubits, and the seal qubits can be in 
any of the diagonal basis eigenstates (i.e., those of the Pauli $\hat{X}$ operator). 
For simplicity, we assume that the eavesdropping
attacker has knowledge of $f$, though in reality he can be worse off.
Similarly, the state encoding for a one 1 bit
is $\rho_1 = \left(^{n}C_{fn}2^{(1-f)n}\right)^{-1}(|1\rangle\langle1|\otimes
|1\rangle\langle1|\cdots|1\rangle\langle1|\otimes|s_1\rangle\langle s_1|
\otimes\cdots\otimes|s_{(1-f)n}\rangle\langle s_{(1-f)n}| + \cdots)$.

It follows that $(\rho_0 - \rho_1)^{\dag} = (\rho_0 - \rho_1) =
\left(^{n}C_{fn}2^{(1-f)n}\right)^{-1}
(|000\cdots\rangle\langle000\cdots| - |111\cdots\rangle\langle111\cdots|)
\otimes\hat{I}\otimes\cdots\otimes\hat{I}
+ \cdots)$, where the summation runs over all $^{n}C_{fn}$ combinations for interspersing
the ${(1-f)n}$ copies of $\hat{I}$ (single-qubit identity) operators amidst the $n$ qubit slots.
After some manipulation, we find:
\be
\label{dhs0}
d^2_{\rm HS} = 
2\cdot{^nC_{fn}}\left[\frac{\sum_{j=0}^{fn-1} [{^{(1-f)n}C_{j}}][{^{fn}C_{j}}]2^{[(1-f)n-j]}}
{\left[^nC_{fn}2^{(1-f)n}\right]^2} \right] < 2^{-[(1-f)n-1]}.
\ee
where we arrive at the inequality noting that $^{n}C_{m}={^{n-m}C_{m}} + 
\sum_{j=0}^{m-1} [{^{n-m}C_{j}}][{^{m}C_{j}}]$, $2m \le n$.
It follows from Eq. (\ref{dhs0}) that the two states
approach indistinguishability exponentially fast in $n$. Thus, no matter what POVM
strategy the attacker chooses, we can increase $n$ to make the chance of detection via
a coherent attack 
arbitrarily small. Further, the layer of secret sharing means that
the chance of launching such an attack and reading the message without being caught
is further exponentially diminished.

\paragraph{Extension to sealing of quantum data:}
The method given above works for sealing classical data. By combining it with quantum
encryption \cite{qcrypt,vwani},
it can be used to implement quantum seals for sealing quantum data.
Quantum encryption works as follows: suppose we have a
$n$-qubit quantum state $|\psi\rangle$ and random sequence $K$ of $2n$
classical bits. Each sequential pair of classical bit is associated with a qubit
and determines which transformation $\hat{\sigma} \in \{\hat{I}, \hat{\sigma}_x,
\hat{\sigma}_y, \hat{\sigma}_z\}$ is applied to the respective qubit. If the
pair is 00, $\hat{I}$ is applied, if it is $01$, $\hat{\sigma}_x$ is applied,
and so on. To one not knowing $K$,
the resulting $|\tilde{\psi}\rangle$ is a complete mixture and
no information can be extracted out of it because the encryption leaves
any pure state in a maximally mixed state, that is:
$(1/4)(\hat{I}|S\rangle\langle S|\hat{I} +
\hat{\sigma}_x|S\rangle\langle S|\hat{\sigma}_x +
\hat{\sigma}_y|S\rangle\langle S|\hat{\sigma}_y +
\hat{\sigma}_z|S\rangle\langle S|\hat{\sigma}_z) = (1/2)\hat{I}$.
However, with knowledge of $K$
the sequence of operations can be reversed
and $|\psi\rangle$ recovered. Therefore, classical data can be used
to encrypt  quantum data.

To seal quantum data $|\psi\rangle$, we proceed as follows:
(1) encrypt  $|\psi\rangle$ using classical data $K$ to  $|\tilde{\psi}\rangle$;
(2) seal classical data $K$ in qubits $|\xi_j\rangle$; (3) intersperse the qubits
$|\xi_j\rangle$ amidst those of $|\tilde{\psi}\rangle$ according to some
combination $C$; (4) seal $C$ using qubits $|\eta_k\rangle$. The total quantum seal
for quantum data consists of the triple $\{|\tilde{\psi}\rangle$, 
$\bigotimes_j |\xi_j\rangle$, $\bigotimes_k |\eta_k\rangle\}$.
We note if the second layer of sealing were absent, and the
quantum seal consisted only of $\{|\tilde{\psi}\rangle$, $\bigotimes_j |\xi_j\rangle\}$,
a malevolent intruder could modify the encrypted data $|\tilde{\psi}\rangle$
without Bob or the verifiers being able to detect it.

In order to read the state $|\psi\rangle$, the reader must first break the 
first layer seal to retrieve $C$, from which he obtains positional information
of the qubits sealing the data $K$. He retrieves $K$ by breaking the seal of
positionally marked qubits. He decrypts $|\tilde{\psi}\rangle$ using $K$ to
obtain $|\psi\rangle$. There are an exponentially large number of ways of
interpolating the $K$-sealing qubits amidst those of $|\tilde{\psi}\rangle$.
Hence the potential attacker must first obtain positional information by breaking
the first layer seal, which will lead to detection with high probability.

\end{document}